\documentclass[twocolumn,prd,preprintnumbers,superscriptaddress,landscape,nofootinbib]{revtex4}
\usepackage[utf8]{inputenc}
\usepackage{amsmath}
\usepackage{amsthm}
\usepackage{amssymb}
\usepackage{float}
\usepackage{braket}
\usepackage{graphicx}
\usepackage{url}
\usepackage{here}
\usepackage{natbib}
\usepackage{rotating}
\usepackage{comment}
\usepackage[colorlinks=true, pdfstartview=FitV, linkcolor=red, citecolor=blue, urlcolor=blue]{hyperref}
\usepackage{slashed}
\usepackage{multirow}
\usepackage[normalem]{ulem}
\graphicspath{{./figures/}}

\newcommand{\be}{\begin{equation}}      
\newcommand{\ee}{\end{equation}}      
\newcommand{\bea}{\begin{eqnarray}}      
\newcommand{\eea}{\end{eqnarray}}

\newcommand{\Tr}{\mathrm{Tr}}

\newcommand{\ctext}[1]{\raise0.2ex\hbox{\textcircled{\scriptsize{#1}}}}
\usepackage{tikz}
\usetikzlibrary{quantikz2}

\theoremstyle{definition}

\theoremstyle{remark}

\usepackage[normalem]{ulem}
\begin{document}
\preprint{}
\title{Exploring Kondo effect by quantum energy teleportation} 
\author{Kazuki Ikeda}
\email{kazuki.ikeda@stonybrook.edu}
\affiliation{Co-design Center for Quantum Advantage, Department of Physics and Astronomy, Stony Brook University, Stony Brook, New York 11794-3800, USA}
\affiliation{Center for Nuclear Theory, Department of Physics and Astronomy, Stony Brook University, Stony Brook, New York 11794-3800, USA}
\author{Rajeev Singh}
\email{rajeevofficial24@gmail.com}
\affiliation{Center for Nuclear Theory, Department of Physics and Astronomy, Stony Brook University, Stony Brook, New York 11794-3800, USA}
\affiliation{C.N. Yang Institute for Theoretical Physics, Stony Brook University, Stony Brook, New York, 11794-3840, USA}
\author{Robert-Jan Slager}
\email{rjs269@cam.ac.uk}
\affiliation{Department of Physics, TCM, University of Cambridge, 19 JJ Thomson Avenue, CB3 0HE, Cambridge, UK}
	\date{\today} 
	\bigskip
\begin{abstract}
We consider a quantum energy teleportation (QET) method to replicate the phase diagram of a one-dimensional $XXZ$ spin chain featuring a Kondo effect coupling. In this setup, the energy supplier and receiver are spatially separated from the point impurity and do not interact directly with it. Nonetheless, they may successfully generate phase diagrams that closely mirror those produced via exact diagonalization. This can be achieved  using only local operations on their respective subsystems, supplemented by classical communication. This feat is made possible due to a critical connection between the energy obtained through the QET approach and the system's quantum entanglement entropy. To substantiate these findings, we initially demonstrate that the quantum entanglement entropy serves as the relevant order parameter for the system. Intriguingly, changes in the gap spacing of the entanglement spectra align with the locations of peaks in both entanglement entropy and energy, as determined by QET. We hypothesize that this theoretical framework could, for example, be validated experimentally using a one-dimensional chain of Rydberg atoms.
\end{abstract}
\maketitle
\section{Introduction}
\label{sec:intro}
The study of quantum many-body systems through local operations has gained significant attention, especially in the context of quantum computation and quantum information theory~\cite{Islam:2015mom,Huang:2020tih,ayral2023quantum}.
In quantum computation, a large system is implemented as a set of qubits, and the physical properties of the system are determined as a binary bit string obtained by measuring each qubit. Quantum computers have the potential to provide a significant advantage in simulating quantum many-body systems, as they can naturally capture  quantum behavior that is inherent to these systems. However, the difficulty in simulating large-scale quantum many-body systems on a quantum computer gives rise to considerable challenges~\cite{Beck:2023xhh}. Quantum systems are sensitive to noise and errors, and maintaining the coherence of qubits over a simulation's duration becomes increasingly challenging as the number of qubits grows. Simulating large quantum systems requires a large number of qubits and a high degree of gate fidelity. As of now, the number of qubits and the level of control needed for accurately simulating large scale quantum many-body systems might exceed the capabilities of current quantum hardware~\cite{Beck:2023xhh}.
One may wonder, whether is it possible to obtain global information on quantum many-body systems using limited quantum resources. In some previous works by one of the authors, this question was addressed by applying a QET strategy to relativistic quantum field theory and symmetry-protected topological phases~\cite{2023arXiv230111712I,2023arXiv230209630I}. Within our context, QET entails a protocol for extracting energy from local systems using only local operations and classical communications (LOCCs). The quantum entanglement in the ground state of the system and the existence of quantum zero-point oscillations of energy are intrinsically important for QET. In addition to the aforementioned analysis of phase transitions, QET has been applied to quantum networks and quantum crypto-systems~\cite{2023arXiv230111884I,Ikeda:2023yhm}. The methods used in the series of studies have been implemented using IBM's real quantum computers which is publicly available, see~\cite{2023arXiv230102666I,Ikeda_Quantum_Energy_Teleportation_2023}. As a subsequent study, the central question we would like to address here is the following: Is it possible to know the presence or absence of a point defect/impurity at a point far from the observer without directly observing the point defect? 

Defects and impurities play a crucial role in shaping the physical properties of materials and systems~\cite{Alloul:2009zz, defect1,defcts2,defect3,PhysRevB.90.241403,ran2009one}. They can significantly impact various material characteristics and behaviors, often leading to unique effects. 
In this work, we turn to the Kondo effect~\cite{kondo1964resistance,Cronenwett_1998}, which is a phenomenon in condensed matter physics, that arises when a magnetic impurity is introduced into a conducting material. In simple terms, the Kondo effect can be explained as follows. A magnetic impurity typically has an unpaired electron spin, which creates a local magnetic moment. This spin interacts with the spins of the conducting electrons in the surrounding material. At higher temperatures when thermal fluctuations dominate, the interaction between the impurity spin and the conducting electrons is weak. The conducting electrons scatter off the impurity without significantly affecting its spin orientation. However, as the temperature is lowered, quantum mechanical effects become more pronounced. At sufficiently low temperatures, the conducting electrons near the impurity are able to form bound states with the impurity spin. These bound states involve complex quantum entanglement between the impurity spin and the conducting electrons. The formation of these bound states has an intriguing consequence. The conducting electrons ``screen'' the impurity spin effectively shielding it from the external magnetic field. This screening leads to a reduction in the local magnetic moment of the impurity. The presence of the impurity spin and the associated electron scattering contribute to an increase in the electrical resistivity of the material at low temperatures. This increase is opposite to the usual behavior in metals, where resistivity typically decreases with decreasing temperature due to the decrease in electron scattering~\cite{kondo1964resistance,Cronenwett_1998}. 

The Kondo effect has been experimentally observed and has other far-reaching implications, from understanding the behavior of strongly correlated electron systems to applications in quantum computing and spintronics~\cite{Cronenwett_1998,Neupane_2015,doi1002/9783527681594,Neupane_2013,B_ri_2012,Buccheri_2016,Dzsaber_2017}. It also provides insights into how quantum entanglement and many-body interactions can lead to emergent behavior in condensed matter systems. There has been renewed interest in the Kondo effect within one-dimensional strongly correlated systems~\cite{PhysRevB.100.165110}. In interacting one-dimensional systems characterized by the Tomonaga-Luttinger universality class~\cite{tomonaga1950remarks,luttinger1963exactly,lieb1965exact,haldane1981luttinger}, the influence of a static impurity potential is significantly altered, being renormalized to either infinite or zero values based on the attractive or repulsive nature of interactions. This peculiar response of Tomonaga-Luttinger liquids to static impurities has garnered significant attention~\cite{Furusaki_2005,lee2016tomonagaluttinger}, prompting further investigations into the impact of dynamic impurities, often involving magnetic impurities in these liquids. Many researchers have examined variations of the generalized Hubbard model that incorporate an impurity spin (with spin value $1/2$)~\cite{HOU1999189,PhysRevB.56.300,z_2019}. Notably, the Kondo temperature, a key energy scale governing the screening of an impurity spin by host electrons, exhibits a power-law relationship with the Kondo exchange coupling. The characteristics of low-energy fixed points have been explored through perturbative renormalization-group analysis and the boundary conformal field theory approach~\cite{Furusaki_1998,Xavier_2022,PhysRevD.99.014040}. 

The rest of the paper is organized as follows. In Sec.~\ref{sec:QET} we explain the theory of QET followed by Sec.~\ref{sec:spin_chain_Kondo}, where we define our model. We provide our results in Sec.~\ref{sec:detecting_phase_transition}, where the phase diagram is reproduced by QET. We also perform a numerical study of the phase transition using the entanglement spectrum. Before concluding, we address the experimental feasibility in Sec.~\ref{sec:exp_feasibility} followed by the conclusion in Sec.~\ref{sec:conclu}.

\begin{figure*}
    \centering
    \includegraphics[width=\linewidth]{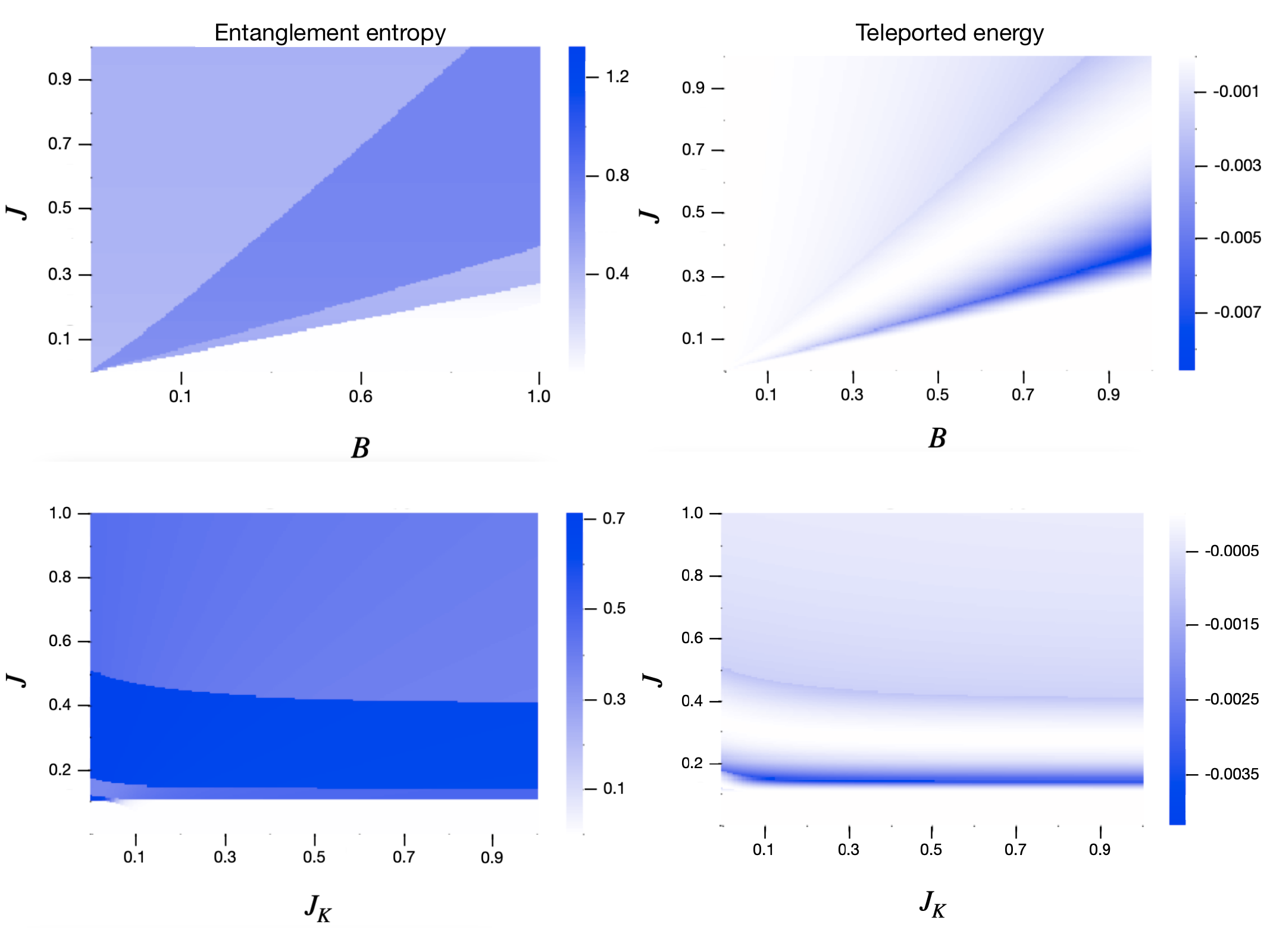}
    \caption{The phase diagram of the $N=8$ Kondo system. Alice at $n_A=4$ has no interaction with the impurity at $n=1$ which is interacting with the nearest site $n=2$. She also has no interaction with Bob at $n_B=7$. The left two panels (upper and lower) correspond to the entanglement entropy and the right two panels (upper and lower) correspond to the teleported energy to Bob's local system. We keep $J_K=0.2$ for upper panels and $B=0.4$ for lower panels.}
    \label{fig:enter-label}
\end{figure*}
\section{Quantum Energy Teleportation}
\label{sec:QET}
In this section, we elucidate the standard procedure of the QET protocol. We note that references to quantum circuit implementations for a variety of configurations can be found in~\cite{2023arXiv230102666I,2023arXiv230111884I,2023arXiv230111712I,Ikeda:2023yhm}.
We begin with the premises that one can express the Hamiltonian as $H=\sum_{n=0}^{N-1}H_n$, where each $H_n$ represents the local Hamiltonian that encompasses the interactions of the nearest neighbors and complies with the constraint
\begin{align}
\begin{aligned}
\label{eq:condition}
\bra{g}H\ket{g}=\bra{g}H_n\ket{g}=0,~\forall \,n\in \{0,1,\cdots,N-1\}\,.
\end{aligned}    
\end{align}
In the above, $\ket{g}$ signifies the ground state of the overall Hamiltonian $H$, it's crucial to recognize that $\ket{g}$ is not invariably the ground state of the local Hamiltonian $H_n$. Engaging in non-trivial local operations, like measuring the ground state, may lead to excited states and augment the energy expectation value, an effect facilitated by the experimental arrangement. Furthermore, $\ket{g}$ is commonly an entangled state.

Here we describe the QET protocol in a spin chain consisting of spin 1/2 particles. We note that the outlined two level setup in principle can be extended to larger on-site Hilbert spaces~\cite{2008PhLA..372.5671H}. To begin, Alice performs a measurement of her Pauli operator $\sigma_{n_A}$ utilizing $P_{n_A}(\mu)=(1+\mu \, \sigma_{n_A})/2$, resulting in either $\mu=-1$ or $\mu=+1$. When measuring the quantum state at subsystem $A$, it disturbs the ground state entanglement and introduces energy $E_A$ from the measurement apparatus into the entire system. In the beginning, the imparted energy $E_A$ is localized around subsystem $A$, but Alice is unable to retrieve it from the system merely through her manipulations at $n_A$. This phenomenon transpires as the entanglement that existed before the measurement ensures that information concerning $E_A$ is preserved in distant parts of the system. To remove the energy from a location other than $n_A$, the quantum many-body characteristic of the system may be employed through LOCC, a process that is explained next.

Through a classical channel, Alice transmits the result of her measurement $\mu$ to Bob, who then implements an operation $U_{n_B}(\mu)$ on his qubit and separately measures his local Pauli operators $X_{n_B}$, $Y_{n_B}$, and $Z_{n_B}$. Following Bob's application of $U_{n_B}(\mu)$ to $P_{n_A}(\mu)\ket{g}$, the resulting density matrix is denoted as $\rho_\text{QET}$, where $\rho_\text{QET}$ is
\begin{equation}
    \rho_{\rm QET}=\sum_{\mu\in\{-1,1\}}U_{n_B}(\mu)P_{n_A}(\mu)\ket{g}\bra{g}P_{n_A}(\mu)U^\dagger_{n_B}(\mu). 
    \label{eq:rho_QET}
\end{equation}
With the utilization of $\rho_{\rm QET}$, the anticipated local energy at Bob's localized system is calculated as $\langle E_{n_B}\rangle=\Tr[\rho_\text{QET}H_{n_B}]$, a value that is typically negative. Owing to the principle of energy conservation, $E_B=-\langle E_{n_B}\rangle (>0)$ is withdrawn from the system by the apparatus conducting the operation $U_{n_B}(\mu)$~\cite{PhysRevD.78.045006}. Through this process, Alice and Bob are able to convey the energy of the quantum system solely by performing operations on their individual local systems.

What's most crucial is that, the technique for measuring $\langle E_{n_B}\rangle$ is well-founded and has been authenticated through actual quantum hardware. Stated differently, the findings and conclusions of this paper do not engage with the question of whether the positive net energy can be retrieved from the system.

In the subsequent, we provide specifics regarding the operations performed by Alice and Bob. The definition of $U_{n_B}(\mu)$ is given by
\begin{equation}
    U_{n_B}(\mu)=\cos\theta I-i\mu\sin\theta\sigma_{n_B},
\end{equation}
where $\theta$ obeys the relations
\begin{align}
    \cos(2\theta)=\frac{\xi}{\sqrt{\xi^2+\eta^2}}\,, \quad
    \sin(2\theta)=-\frac{\eta}{\sqrt{\xi^2+\eta^2}}\,,
\end{align}
with
\begin{align}
\xi=\bra{g}\sigma_{n_B}H\sigma_{n_B}\ket{g}\,, &\qquad \eta=\bra{g}\sigma_{n_A}\dot{\sigma}_{n_B}\ket{g}\,,\nonumber\\ \dot{\sigma}_{n_B}&=i[H,\sigma_{n_B}]\,.
\end{align}
The local Hamiltonian must be selected in such a way that $[H,\sigma_{n_B}]=[H_{n_B},\sigma_{n_B}]$. After Bob applies the operation $U_{n_{B}}(\mu)$ to $P_{n_A}(\mu)\ket{g}$, the average quantum state $\rho_\text{QET}$ is achieved. Subsequently, the average energy that Bob measures is
\begin{equation}
\label{eq:QET}
    \langle E_{n_B}\rangle=\Tr\left[\rho_\text{QET}H_{n_B}\right]=\frac{1}{2}\left[\xi-\sqrt{\xi^2+\eta^2}\right]\,.
\end{equation}
The value is negative when $\eta\neq 0$. It is anticipated that if there is no energy loss, the positive energy of $-\langle E_{n_B}\rangle$ is conveyed to Bob's device following the measurement, in accordance with the principle of energy conservation.
\begin{figure*}
    \centering
\includegraphics[width=\linewidth]{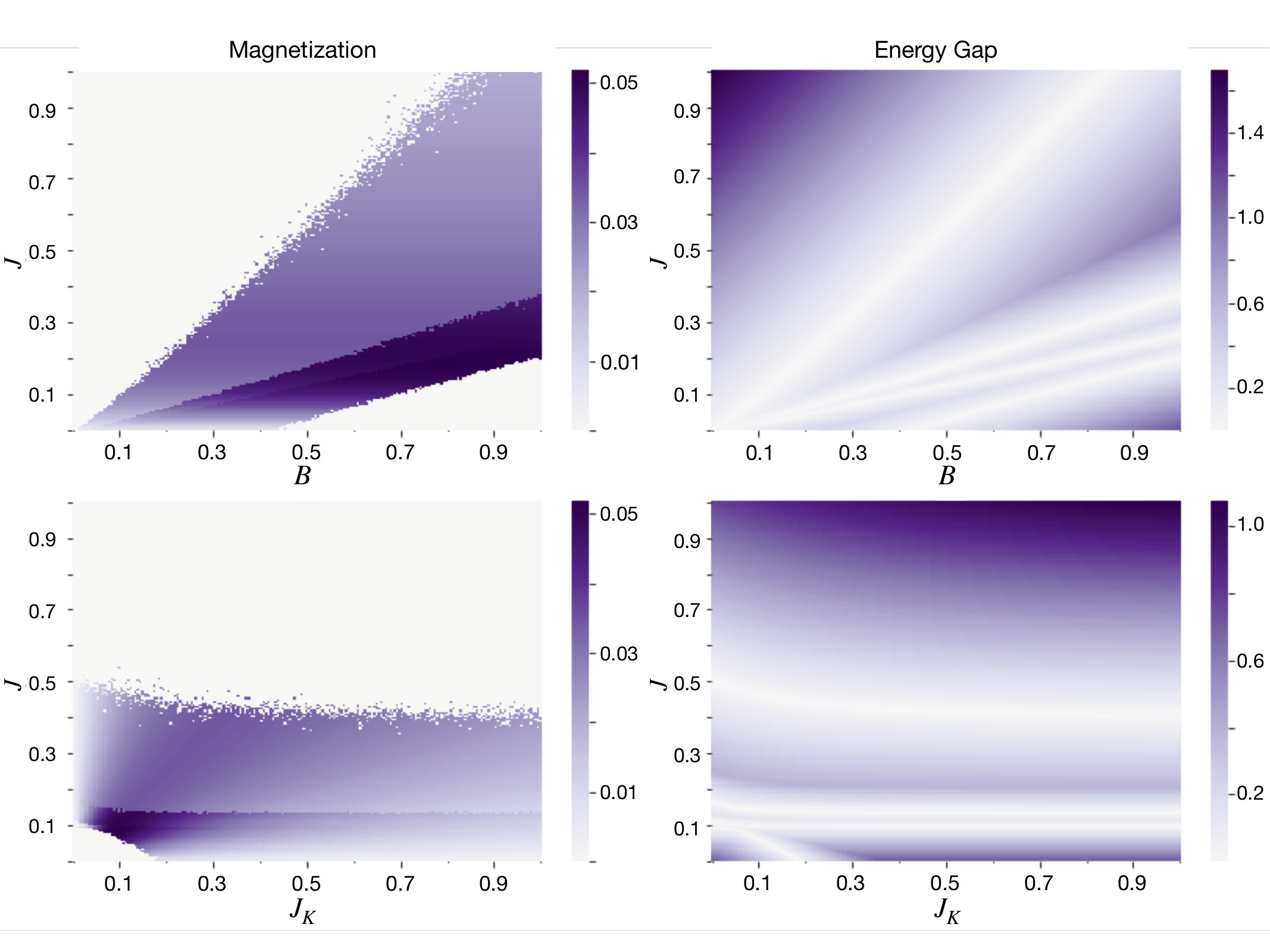}
    \caption{Magnetic field (left panel, upper and lower) and the energy gap (right panel, upper and lower). We keep $J_K=0.2$ for upper panels and $B=0.4$ for lower panels.}
    \label{fig:mag}
\end{figure*}
\section{Spin chains of Kondo effect}
\label{sec:spin_chain_Kondo}
In this section, we would like to explore the Kondo effect through QET and therefore, we examine the subsequent $XXZ$ spin chain model~\cite{laflorencie2008kondo,PhysRevB.58.5529}
\begin{eqnarray}\label{eq:Ham}
    H&=&J\sum_{i}(S^x_iS^x_{i+1}+S^y_iS^y_{i+1}+\Delta S^z_iS^z_{i+1})\nonumber\\    
    &+& J_K(S^x_1S^x_{n}+S^y_1S^y_{n}+\Delta S^z_1S^z_{n})+B\sum_iS^z_i\,.
    \label{eq:Kondo_hamiltonian}
\end{eqnarray}
The term $J\sum_{i}(S^x_iS^x_{i+1}+S^y_iS^y_{i+1}+\Delta S^z_iS^z_{i+1})$ describes the interaction between neighboring spins $S_i$ in the host material, with $\Delta$ allowing for an anisotropic interaction and $J_K(S^x_1S^x_{n}+S^y_1S^y_{n}+\Delta S^z_1S^z_{n})$ is the Kondo term, which couples, through coupling $J_K$, the impurity spin $S_1$ with a spin $S_n$ in the host material.
$B\sum_iS^z_i$ represents the influence of an external magnetic field $B$ on all the spins in the system. Here, $n$ is an indicative of the position of an impurity.

We first recall the phase diagram of the model as shown in Fig.~\ref{fig:enter-label}.  
In the Kondo system phase diagram with system size $N=8$, see Fig.~\ref{fig:enter-label}, Alice, located at $n_A=4$, does not interact with the point impurity at $n=1$, which itself is in interaction with the adjacent site at $n=2$. Additionally, Alice has no interaction with Bob, who is at $n_B=7$. The left-hand side panels, both upper and lower, display the entanglement entropy, while the right-hand side panels, also upper and lower, illustrate the energy teleported to Bob's local system. The parameters utilized are $J_K=0.2$ for upper panels and $B=0.4$ for lower panels. We observe that both entanglement entropy and teleported energy agree and share similar features.

The magnetic field profile used for our analysis is 
\begin{equation}
m=\frac{1}{N} \bra{\psi}\sum_{i=1}^NZ_i\ket{\psi}\,, 
\end{equation}
which is shown in the left panel plots (upper and lower) in Fig.~\ref{fig:mag}. Crucially, we observe the correspondence between the teleported energy of Fig.~\ref{fig:enter-label} and energy gap of Fig.~\ref{fig:mag}.
\section{Detecting phase transition through QET}
\label{sec:detecting_phase_transition}
An intriguing connection exists between the entanglement entropy and the energy teleported.
The variation in entanglement entropy before and after Alice's measurement can be computed as follows
\begin{align}
\label{eq:e}
    \Delta S_{AB}&=S_{AB}-\sum_{\mu\in\{\pm1\}}p_\mu S_{AB}(\mu)\,.
\end{align}
The first term on the right hand side of the equation, $S_{AB}$, depicts the entanglement entropy between Alice and Bob prior to her measurement, whereas, in the second term, $S_{AB}(\mu)$ is the entanglement entropy after Alice's measurement and probability distribution of $\mu$ is represented by $p_\mu$.

Within the context of the minimal model $H=h(Z_0+Z_1)+2kX_0X_1+\text{const.}$~\cite{2011arXiv1101.3954H}, we know that the variation of the entanglement entropy satisfies
\begin{equation}
    \Delta S_{AB}\ge -\frac{1+\sin^2\lambda}{2\cos^3\lambda}\ln\frac{1+\cos\lambda}{1-\cos\lambda}\frac{E_B}{\sqrt{h^2+k^2}}\,,
\end{equation}
with $\lambda=\arctan\left(k/h\right)$ implying that entanglement is expended to facilitate energy teleportation, a consequence of Bob carrying out conditional operations based on Alice's measurement outcomes. In this regard, an analog of  Maxwell's daemons circumstance can be obtained~\cite{PhysRevA.56.3374,2015NatCo...6.7498I}. While it remains uncertain whether this relationship is universally applicable, similar outcomes have at least been confirmed numerically~\cite{2023arXiv230111712I}. Additionally, it is established that a specific relationship strictly holds
\begin{equation}
    S_{A\bar{A}}\ge\frac{E^2_B}{4\|H_B\|^2}\,,
\end{equation}
where $S_{A\bar{A}}$ denotes the entanglement entropy between the subsystem $A$ and its complement $\bar{A}$. 
In the denominator, $H_B$ represent the local Hamiltonian which allows to extract energy $E_B$~\cite{PhysRevA.87.032313}. This will be studied extensively in the future publication~\cite{IkedaLowe}.
To investigate the relation between QET and entanglement entropy on a deeper level, we also examined into the role of entanglement entropy as an order parameter of the phase transition. We use the entanglement entropy $S_{LR}$ between the left and right sides of the system. The results are shown in Figs.~\ref{fig:enter-label} and~\ref{fig:mag}. 

\begin{figure}
\includegraphics[width=\linewidth]{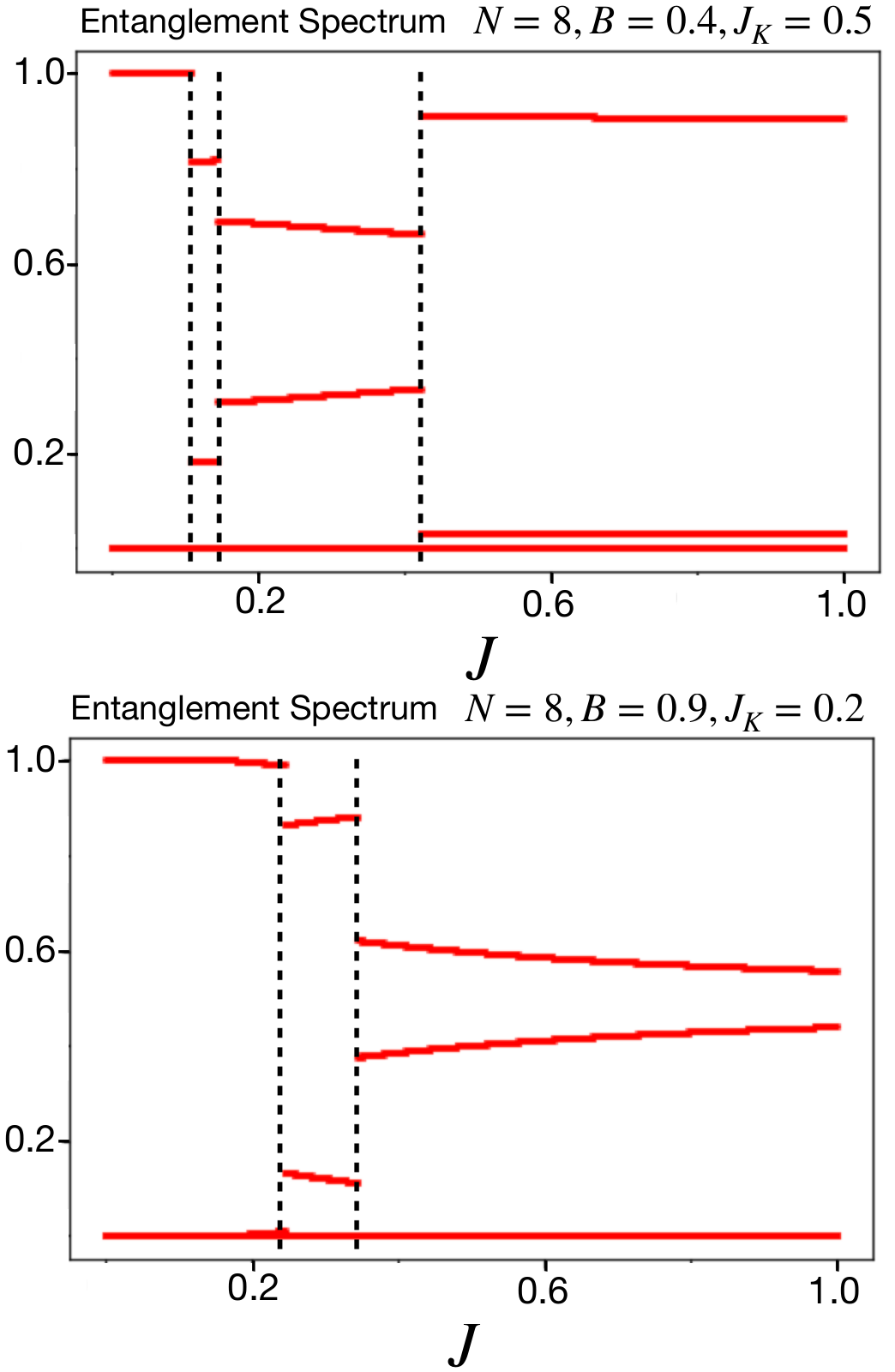}
    \caption{Entanglement spectrum for the Kondo chain as obtained by the Gram--Schmidt process after dividing the system.  }
    \label{fig:ES}
\end{figure}

Additionally, in the following we investigate the entanglement spectrum, see Fig.~\ref{fig:ES}. Let's first assume that the dimension of the Hilbert space of subsystem $L$
is not greater than that of subsystem $R$: $\text{dim}\mathcal{H}_L\le\text{dim}\mathcal{H}_R$. The Gram--Schmidt process for orthonormalizing a quantum state is given as
\begin{equation}
\ket{\psi} = \sum_{\alpha=1}^{\text{dim}\mathcal{H}_L}\xi_\alpha\ket{\psi^L_\alpha}_L\ket{\psi^R_\alpha}_R\,,
\end{equation}
and the coefficients are denoted by $$\xi_\alpha\,, \quad \rm{where} \quad \alpha=1,\cdots,\text{dim} \mathcal{H}_L\,.$$ 
These coefficients then make up the entanglement spectrum~\cite{PhysRevB.81.064439,Ikeda:2023zil} depicted in Fig.~\ref{fig:ES} where the upper panel is for the parameters, $N=8,\,B=0.4,\,J_K=0.5$, whereas, for the lower panel we have $N=8,\,B=0.9,\,J_K=0.2$. We observe that the entanglement spectrum is enhanced for lower value of magnetic field and higher value of Kondo coupling.
\section{Experimental feasibility}
\label{sec:exp_feasibility}
While Kondo physics arises in many experimentally relevant platforms, a rather direct route to implementing the described QET scheme arises in the context of cold atoms systems, see e.g.~\cite{Cronenwett_1998,Neupane_2015,doi1002/9783527681594,Neupane_2013,B_ri_2012,Buccheri_2016,Dzsaber_2017}. A particular appealing platform in this regard entails Rydberg atoms. In such systems one considers (transitions to) highly excited states in e.g. Rubidium atoms~\cite{rydberg1,rydberg2, rydberg3}. As such, these electrons in these large orbits can be well manipulated with light-matter interactions, but more importantly have an intrinsically large dipole moment that connect neighboring atoms.
As a result of the mutual interactions within an ensemble of Rydberg atoms, only one atom can be excited as the interaction ensures the other levels get off resonance. In particular within a volume, set by a critical distance, $R_b$ any two atoms cannot be excited at the same time. This Rydberg blockade enables to incorporate effective interactions and can, for example, be used to simulate topological spin liquids~\cite{rydber4}.
In the limiting case where the dipole interactions are much smaller than the difference in energy levels of the atoms involved, and thus can be treated perturbatively, the effective Rydberg blockade interaction terms is proportional to $R^{-6}$ in terms of the inter-atomic distance $R$.

\begin{figure}
    \centering
    \includegraphics[width=\linewidth]{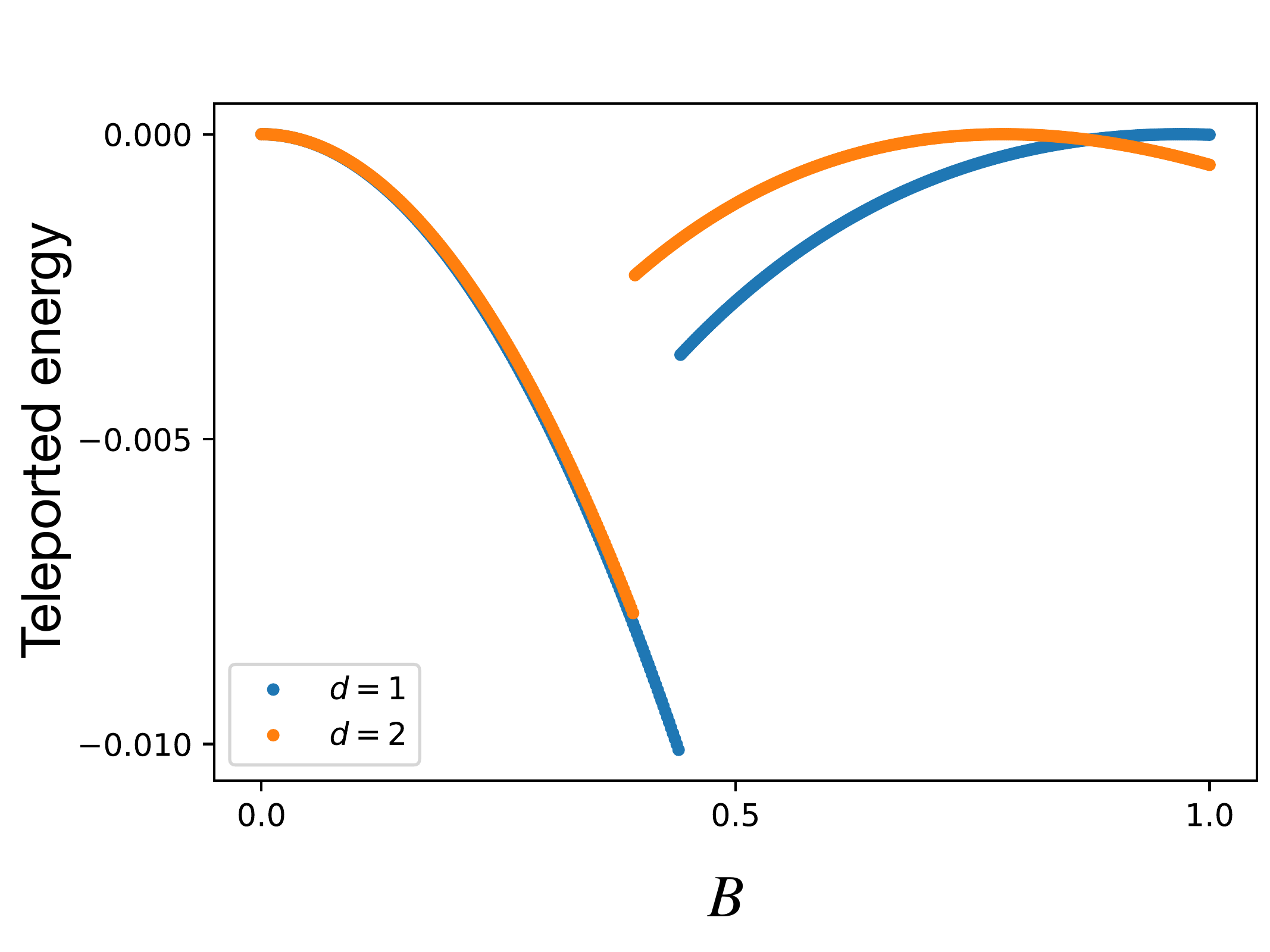}
    \caption{The plot shows the dependence of teleported energy at Bob's location $(y-axis)$ on the magnetic field $B$ $(x-axis)$ with different position of the impurity coupling in the Rydberg-like potential: $J_{K,d=2}/J_{K,d=1}=1/d^6$, where $d$ is the distance from the impurity. We used the following parameters $N=8, n_A=5, n_B=8, n=1, J=0.5, J_{K,d=1}=0.2$ for our analysis.}
    \label{fig:Rydberg}
\end{figure}

To investigate the compatibility with an effective $1/R^{6}$ coupling, as in Refs.~\cite{rydberg1,rydber4,PhysRevX.11.031005}, we study the dependence of our model, Eq.~\eqref{eq:Kondo_hamiltonian}, on the distance $d$ from the impurity, that is shown in Fig.~\ref{fig:Rydberg}. Here we put the impurity at $n=1$ that is coupled with an atom at $n=2~(d=1)$ or $n=3~(d=2)$ with the coupling constant $J_K$. In both the cases, Alice $(n_A=5)$ and Bob $(n_B=8)$ are away from these atoms. As is evident from Eq.~\ref{eq:Ham}, the effect of Rydberg atoms may become significant when $d=2$.
We observe, from Fig.~\ref{fig:Rydberg} that the teleported energy at Bob's location for constant $J_K$ overlaps for $d=1$ and $d=2$ at lower value of magnetic field $B$ but differs at higher values of $B$. We however observe that for our model, Eq.~\eqref{eq:Kondo_hamiltonian}, the energy teleportation does not depend on how far an atom is from the impurity.
We also note that it does not depend on the potential, for example $1/R^6$, and therefore, the energy will be teleported. Indeed, we have corroborated these statements also numerically within the same setup for general larger separations $n>3$.
This result implies that even without impurity information, Alice and Bob can know the impurity effects and phase diagram through QET.
\section{Conclusion and discussion}
\label{sec:conclu}
In conclusion, our study has utilized a recent quantum energy teleportation strategy to faithfully reproduce the phase diagram of a one-dimensional spin chain governed by Kondo effect coupling. Remarkably, this was accomplished even though the entities supplying and receiving energy are located at a spatial distance from the point impurity, without any direct interaction with it. By leveraging only local operations on their individual subsystems, and provided by classical communication, these entities are able to generate phase diagrams that are in direct agreement with those derived from exact diagonalization techniques. The underpinning principle that enables this achievement is an essential correlation between the energy procured via QET and the quantum entanglement entropy of the system. To corroborate this, we initially identified that the quantum entanglement entropy operates as the system's pertinent order parameter. Intriguingly, shifts in the entanglement spectra's gap spacing are found to coincide with peaks in both the quantum entanglement entropy and the energy metrics acquired through QET. We propose that these theoretical insights could be substantiated in an experimental setting, potentially through one-dimensional atomic simulators, such as Rydberg arrays.
\section*{Acknowledgements}
The work of K.I. was supported by the U.S. Department of Energy, Office of Science, National Quantum Information Science Research Centers, Co-design Center for Quantum Advantage (C2QA) under Contract No. DESC0012704.
R.S. is supported by the Polish NAWA Bekker program No. BPN/BEK/2021/1/00342 and Polish NCN Grant No. 2018/30/E/ST2/00432. R.S. also thanks the Institute for Nuclear Theory at the University of Washington for its kind hospitality and stimulating research environment. R.~J.~S  acknowledges funding from Trinity College, University of Cambridge and the Winton programme. This research was supported in part by the INT's U.S. Department of Energy grant No. DE-FG02-00ER41132.
\bibliographystyle{apsrev4-1.bst}
\bibliography{ref}
\end{document}